\documentclass[runningheads]{llncs}

\usepackage{amsmath,amssymb}
\usepackage{graphicx}
\usepackage[caption=false]{subfig}
\usepackage{multirow}
\usepackage{tabulary}
\usepackage{url}  

\begin{document}
\title{ Multi-level Graph Drawing \\ using Infomap Clustering
\thanks{Research supported by ARC Linkage Grant with Oracle labs.}}

\author{Seok-Hee Hong\inst{1} \and
Peter Eades\inst{1} \and
Marnijati Torkel\inst{1} \and
Ziyang Wang\inst{1} \and \\
David Chae\inst{1} \and
Sungpack Hong\inst{2} \and
Daniel Langerenken\inst{2} \and
Hassan Chafi\inst{2}}
\institute{University of Sydney, Australia \\
\email{\{seokhee.hong,peter.eades,mtor0581,zwan0130,min.chae\}@sydney.edu.au} \and
Oracle Research lab, US \\
\email{\{sungpack.hong,daniel.langerenken,hassan.chafi\}@oracle.com}}

\authorrunning{S.H. Hong et al.}

\date{}
\maketitle

\begin{abstract}
{\em Infomap clustering} finds the community structures that minimize the expected description length of a random walk trajectory; algorithms for infomap clustering run fast in practice for large graphs.
In this paper we leverage the effectiveness of Infomap clustering combined with the multi-level graph drawing paradigm.
Experiments show that our new Infomap based multi-level algorithm produces good visualization of large and complex networks, with significant improvement in quality metrics.

\end{abstract}

\section{Introduction}
\label{sec:introduction}

The {\it multi-level graph drawing} is a popular approach to visualize large and complex graphs to improve the quality of
drawings. It recursively coarsens the graph and then uncoarsens the drawing using layout refinement.
There are a number of multi-level graph drawing algorithms available~\cite{gajer2002grip,hachul2005drawing,hu2005efficient,koren2002fast,An,quigley2001fade,walshaw2003multilevel}.
They mainly differ in the {\em coarsening} method.

Clustering is a widely used analysis method for identifying groups with strong similarity, or  communities in the data.
Graph clustering is to partition a graph such that vertices in the same cluster are more interconnected.
{\em Infomap clustering} computes clusters by translating a graph into a map, which decomposes the myriad nodes and links into modules that represent the graph~\cite{Infomap}.
It maximizes an objective function called the minimum description length of a random walk trajectory, where the approximation to the optimal solution can be computed quickly.
Infomap performed the best in community finding experiments~\cite{Infomapexp}.

In this paper, we present a new multi-level graph drawing algorithm based on Infomap clustering.
More specifically, we leverage the effectiveness of Infomap clustering, combined with the multi-level graph drawing paradigm.
Experiments with real-world large and complex networks such as protein-protein interaction networks, Facebook graph, Autonomous Systems (AS) graphs
as well as benchmark graphs show that our new multi-level algorithms produce good visualization with significant improvement in quality metrics, including shape-based metrics~\cite{DBLP:conf/gd/Shape}, edge crossing and stress.
It also requires a small number of coarsening steps for medium to large graphs, which makes it fast to run.

\section{Related Work}
\label{se:related}

Hadany and Harel presented the multi-scale method using an edge contraction based coarsening method and a force-directed layout preserving topological properties such as cluster size and vertex degree~\cite{hadany2001multi}.
Koren and Harel presented FMS, which used  a {\em k}-center approximation based coarsening method
and a force-directed layout with a beautification~\cite{koren2002fast}.

Walshaw presented a multi-level algorithm using a matching, by repeatedly collapsing maximal independent subsets of graph edges, and a grid variant of Fruchterman-Reingold~\cite{fruchterman1991graph} layout~\cite{walshaw2003multilevel}.
Gajer {\em et al.} presented GRIP using a  maximum independent set filtration based coarsening method, and an intelligent initial placement of vertices based on both graph and Euclidean distances~\cite{gajer2002grip}.

Quigley and Eades presented FADE using the quad tree,
and  Barnes-Hut $n$-body method~\cite{barnes1986hierarchical} for approximation of the repulsive force computation in a force-directed layout~\cite{quigley2001fade}.
Hachul and Junger presented FM\textsuperscript{3} using similar method to compute the repulsive forces between vertices,
where subgraphs with small diameter, called solar system, are partitioned and collapsed to obtain a multi-level representation~\cite{hachul2005drawing}.
Hu presented the {\it sfdp} layout, also using the Barnes-Hut approximation method~\cite{hu2005efficient}.
Frishman and Tal~\cite{DBLP:journals/tvcg/FrishmanT07} presented a multi-level force directed graph layout on the GPU, based on spectral partitioning and Kamada-Kawai layout~\cite{kamada1989algorithm}.
Bartel {\em et al.} presented an experimental study for extensive comparison of various multi-level algorithms,
using a combination of coarsening methods, initial placement and graph layout methods~\cite{bartel2011experimental}.

More recently, Meyerhenke {\em et al.} presented a multi-level algorithm using a label propagation method for the coarsening step, and Maxent stress optimisation layout~\cite{DBLP:journals/tvcg/Yifan} on shared memory parallelization~\cite{DBLP:conf/gd/Martin}.
Nguyen and Hong used fast {\it k-core} coarsening method, which can be computed in linear time~\cite{An}.

\section{Infomap based Multi-level Algorithm}
\label{sec:algorithm}

The multi-level graph drawing algorithm is an iterative process consisting of the following three steps:
 {\it coarsening}, {\it initialization} (or {\em placement}), and {\it graph layout} ({\it or refinement}).
Roughly speaking, the coarsening step is to  cluster vertices to define a smaller graph, recursively
until the size of the graph falls below the threshold, resulting in a coarse graph hierarchy, $G_0$, $G_1$, \ldots, $G_L$.
The layout of graph $G_{L}$ is then extended to the layout of graph $G_{L-1}$
by {\em placement} (i.e., add vertices back to the layout) and {\em refinement} step.
Recursively, these steps extend the layout of graph $G_L$ to $G_0$ by repetitively interpolating from $G_i$ to $G_{i-1}$.
In each iteration, the layout of $G_i$ is used to compute an initial placement of $G_{i-1}$,
and then the layout algorithm is applied to refine the layout.


\medskip
\noindent {\bf 3.1 Coarsening: Infomap Clustering}

Let $G=(V, E)$ be a graph with vertex set $V$ and edge set $E$.
The coarsening step computes a graph level hierarchy by iteratively computing a sequence of smaller graphs $G_0$, $G_1$, $G_2$, \ldots, $G_L$, where the original graph $G = G_0$.
At each level, a coarser graph (or clustered graph) is computed by combining a sets of vertices belong to the same cluster in $G_i$ and replacing into a single vertex in $G_{i+1}$, recursively until the predefined stop criterion is satisfied.

{\em Infomap clustering} finds community structure that minimizes the expected description length of a random walk trajectory~\cite{Infomap}.
It computes clusters by translating a graph into a map, which decomposes the myriad nodes and links into modules that represent the graph.
The algorithm maximizes an objective function called the Minimum Description Length.

We first compute the Infomap clustering of $G$, and partition the vertex set $V$ into $V_i$ based on the clusters.
More specifically, we define a clustered graph  with a weighted vertex set (i.e., the number of vertices belong to each cluster) and a weighted edge set (i.e., the number of edges between the partitioned vertex set).
The vertices $u_1, u_2, \ldots, u_k \in V_i$ are merged to form a new cluster vertex $v \in V_{i+1}$, where the weight of $v$ is computed as $|v| = |u_1| + |u_2| + \ldots + |u_k|$. Similarly, the weight of the collapsed edges are computed as the sum of the weights of the edges that it replaces.
This coarsening phase stops when the resulting clustered graph has a small size (say 50) or there is no reduction in terms of size.

\medskip
\noindent{\bf 3.2 Initialization: Placement}

This step aims to compute a good initial layout of $G_{i-1}$ using the layout of $G_i$.
Let $v_i \in V_i$ of $G_i$ corresponds to a cluster of vertices $u_1, u_2, \ldots, u_k \in V_{i-1}$ of $G_{i-1}$.
We add back vertices $u_j, j = 1, \ldots, k$ to the layout of $G_i$ by initializing the positions of $u_j$ using the position of $v_i$.
Here we use the following three variations.

- Circle placement: It places all $u_j, j = 1, \ldots, k$ at the circle with a small radius, where the center of the circle is the location of $v_i$.

- Barycenter placement: It places each vertex at the barycenter of its neighbors~\cite{gajer2002grip}.

- Zero placement: It places all $u_j, j = 1, \ldots, k$ at the same position as $v_i$ with small perturbation~\cite{walshaw2003multilevel}.

\medskip
\noindent {\bf 3.3 Refinement: Force-directed Layout}

The initial layout of $G_{i-1}$ is recursively refined at each level using a force-directed algorithm.
We use layout algorithms, previously used in other multi-level graph drawing algorithm experiments~\cite{bartel2011experimental}:

- FR: Fruchterman and Reingold layout~\cite{fruchterman1991graph}.

-  FRG: grid variant of Fruchterman and Reingold layout, used in~\cite{walshaw2003multilevel}.

- FME (Fast-Multipole Embedder): an improvement of NME (New Multipole Method) layout of FM\textsuperscript{3}~\cite{hachul2005drawing}, designed for a multi-level method in~\cite{bartel2011experimental}.

\section{Experiments}
\label{subsec:experiment-algorithm}


We implemented Infomap clustering based multi-level algorithm using OGDF~\cite{DBLP:reference/crc/OGDF},
which was used in the comparison experiments of multi-level algorithms~\cite{bartel2011experimental}.
We used a standard Dell laptop with Intel Core i7, 16 GB RAM.

We first experimented with three different placement methods, and found that
there is no significant difference in terms of layout quality.
We choose the barycenter placement, which shows slightly better performance,
with three layouts FR, FRG and FME for comparison.

More specifically, we have the following variations for comparison:

- InfomapFR: Infomap multi-level algorithm with FR layout

- InfomapFRG: Infomap multi-level algorithm with FR grid variant layout

- InfomapFME: Infomap multi-level algorithm with FME layout


The experiment was conducted with real-world benchmark data sets including social networks such as facebook,
biological networks such as protein-protein interaction networks,
and benchmark graphs used in previous work~\cite{bartel2011experimental,An,walshaw2003multilevel}.

Table~\ref{table:graph-level-size} shows the details of the data sets, the number of coarsening levels and runtime (seconds), where $D$ represents the density of a graph $G$ and $L$ represents the number of levels.
We can clearly see that the Infomap coarsening method produced small number of levels such as 2 or 3 for most of data sets.
Overall, Infomap clustering runs quite fast for medium size graphs.

\begin{table}
 \begin{center}
  \begin{tabular}{ | l | r | r | r | r | r | r | r | r | r | r | r | r |}
    \hline
    Graph $G$ & $|V_0|$ & $|E_0|$ & $D$ & $L$ & Time & $|V_1|$ & $|E_1|$ & $|V_2|$ & $|E_2|$ & $|V_3|$ & $|E_3|$ \\ \hline
    \emph{G\_15\_0} & 1785 & 20459 & 11.5 & 2 & 0.02 & 59 & 100& 9 & 8 & &\\ \hline
    \emph{nasa1824} & 1824 & 18692 & 10.3& 2 & 0.02 & 53 & 217 & 5 & 7 & &\\ \hline
    \emph{G\_4\_0} & 2075 & 4769 & 2.3 & 2 & 0.02 & 89 & 326 & 8 & 11 & & \\ \hline
    $yeastppi$ & 2361 & 7182 &3.0 & 2 & 0.04 & 302 & 1923 & 101 & 0& & \\ \hline
    $soc\_h$ & 2426 & 11630 & 4.8 & 2 & 0.02 & 301 & 1088 & 149& 1 & & \\ \hline
    $oflights$ & 2939 & 15677 & 5.3 & 2 & 0.03 & 170 & 477 & 19 & 24 & & \\ \hline
    $ecolippi$ & 3796 & 78120 & 20.6 & 2 & 0.03 & 245 & 2453 & 53& 1& & \\ \hline
    $facebook$ & 4039 & 88234 & 21.9 & 2 & 0.02 & 93& 272& 7& 11 & & \\ \hline
    \emph{3elt} & 4720 & 13722 & 2.9 & 2 & 0.05 & 189 & 489& 17 & 35 & & \\ \hline
    $USpowerGrid$ & 4941 & 6594 & 1.3 & 2 & 0.18 & 489& 963& 44& 104& & \\ \hline
    \emph{as19990606} & 5188 & 10974 & 2.1 & 2 & 0.17 & 368& 2034& 12 & 38 & & \\ \hline
    $commanche\_dual$ & 7920 & 19800 & 2.5 & 2 & 0.24 & 503 & 1365 & 34 & 71 & & \\ \hline
    \emph{p2p-Gnutella05} & 8846 & 31839 & 3.6 & 2 & 0.20 & 830 & 18154 & 3 & 0 & & \\ \hline
    \emph{astroph2001} & 16046 & 121251 & 7.6 & 3 & 0.61 & 1219 & 9333 & 395 & 68 & 369& 0\\ \hline
    \emph{condmat2001} & 16264 & 47594 & 2.9 & 3 & 1.33 & 1720 & 4574 & 798& 774& 726& 0\\ \hline
    \emph{crack-dual} & 20141 & 30043 & 1.5 & 3 & 1.16 & 1357 & 3633& 84& 216& 10& 18\\ \hline
    \emph{bcsstk31} & 35588 & 608502 & 17.1 & 2 & 0.36 & 453& 2295& 25& 44& & \\ \hline
    \emph{shock-9} & 36476 & 71290 & 2.0 & 3 & 1.17 & 1351& 3852 & 74& 191& 8& 14\\ \hline
    \emph{del16} & 65536 & 196575 & 3.0 & 3 & 1.95 & 1981 & 5921& 101& 290& 8 & 16 \\ \hline
  \end{tabular}
  \end{center}
 \caption{Data sets, size, number of levels ($L$) and runtime.}
 \label{table:graph-level-size}
\end{table}

\begin{figure}[h!]
  \centering
  \subfloat[Shape-based (larger, better)]{\label{fig:shape1}%
    \centering
     \includegraphics[width=0.35\textwidth]{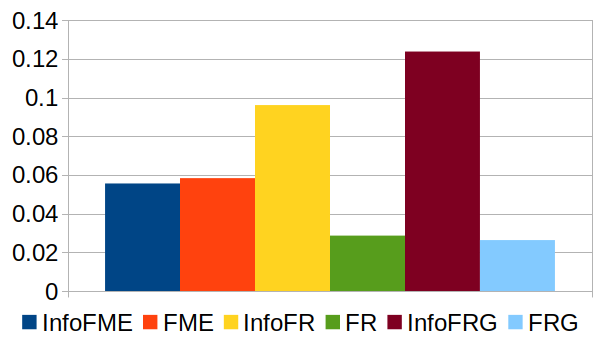}
  }
  \subfloat[stress (smaller, better)]{\label{fig:stress1}%
    \centering
      \includegraphics[width=0.35\textwidth]{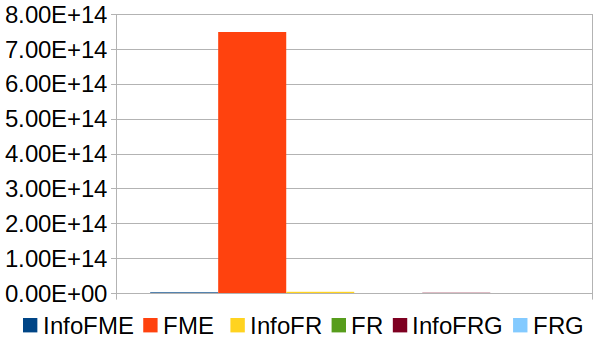}
   }
    \subfloat[crossing (smaller, better)]{\label{fig:cross1}%
    \centering
      \includegraphics[width=0.35\textwidth]{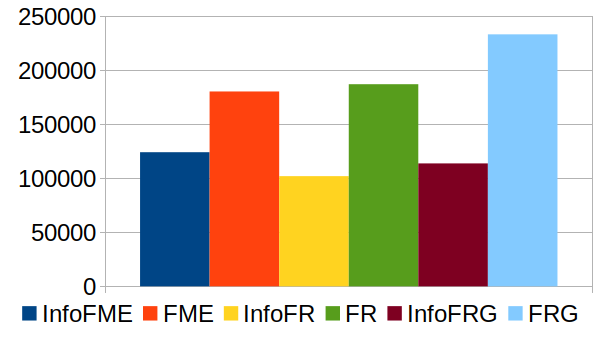}
   }
   \hfill
  \centering
  \subfloat[Shape-based]{\label{fig:shape2}%
    \centering
      \includegraphics[width=0.20\textwidth]{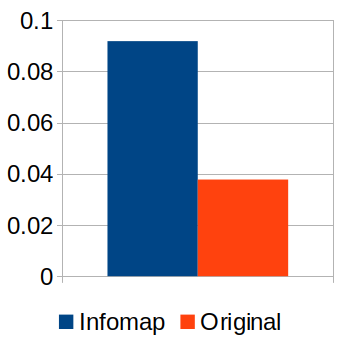}
  }
  \subfloat[stress]{\label{fig:stress2}%
    \centering
      \includegraphics[width=0.20\textwidth]{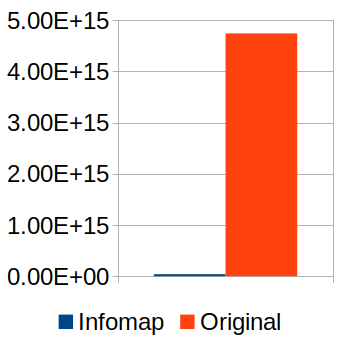}
  }
  \subfloat[crossing]{\label{fig:cross2}%
    \centering
      \includegraphics[width=0.20\textwidth]{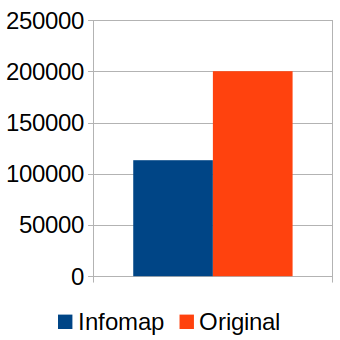}
  }
    \subfloat[Improvement]{\label{fig:shape4}%
    \centering
      \includegraphics[width=0.20\textwidth]{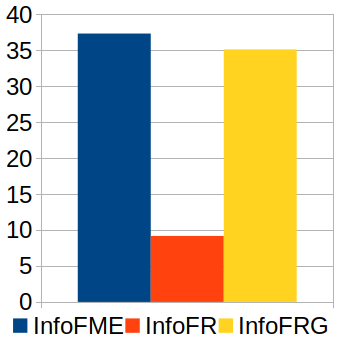}
  }
  \subfloat[Infomap vs. $FM^3$ ]{\label{fig:shape3}%
    \centering
      \includegraphics[width=0.20\textwidth]{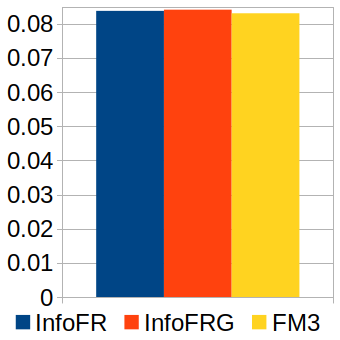}
  }
\caption{Significant improvement in quality metrics: (a)(b)(c) Average of metrics per layout: InfomapFME (blue), FME (red), InfomapFR (yellow), FR (green), InfomapFRG (brown), FRG (cyan);
(d)(e)(f) Average of metrics: Infomap (blue) vs. Original (red);
(g) Average of improvement by Infomap over Original in Shape-based metrics: InfomapFME (blue), InfomapFR (red), InfomapFRG (yellow);
(h) Average of shape-based metrics: InfomapFR (blue), InfomapFRG (red), $FM^3$ (yellow). }
\label{fig:average}
\end{figure}

\noindent{\bf Comparison of Quality Metrics: }
For large and complex graphs, edge crossing may not a suitable metric to measure the quality of drawings~\cite{DBLP:conf/gd/Shape,DBLP:conf/gd/KobourovPS14}.
We used the {\em shape-based metrics}~\cite{DBLP:conf/gd/Shape}; this is a new graph drawing quality measure specially designed for large graphs.
Roughly speaking, the shape-based metrics measure the \emph{faithfulness} of graph drawing,
i.e., how well the {\em shape} of the drawing represents the structure (or shape) of the graph.

Figures~\ref{fig:average}(a), (b) and (c) show the comparison of {\em average metrics} between six layouts
(i.e., Infomap multi-level vs. FME, FRG, FR original layouts)
 using {\em shape-based quality metrics} ($Q$), {\em stress} and {\em edge crossings}.
Clearly, we can see that Infomap multi-level layouts perform significantly better than the original layouts.
In general, InfomapFR and InfomapFRG perform better than InfomapFME.
%
Figures~\ref{fig:average}(d), (e) and (f) show the {\em average metrics} between Infomap multi-level and original layouts.
Overall, we can see that Infomap multi-level layouts outperform original layouts.
Figure~\ref{fig:average}(g) shows the {\em average improvement} by Infomap multi-level layouts over original layouts in shape-based metrics
(i.e., ($Q_{Infomap}/Q_{Original} - 1$)).
Clearly, significant improvement was achieved by InfomapFME and InfomapFRG.



\medskip

\begin{figure}[h!]
  \centering
  \subfloat[FME]{\label{fig:3elt_original_fme}%
    \centering
     \includegraphics[width=0.3\textwidth]{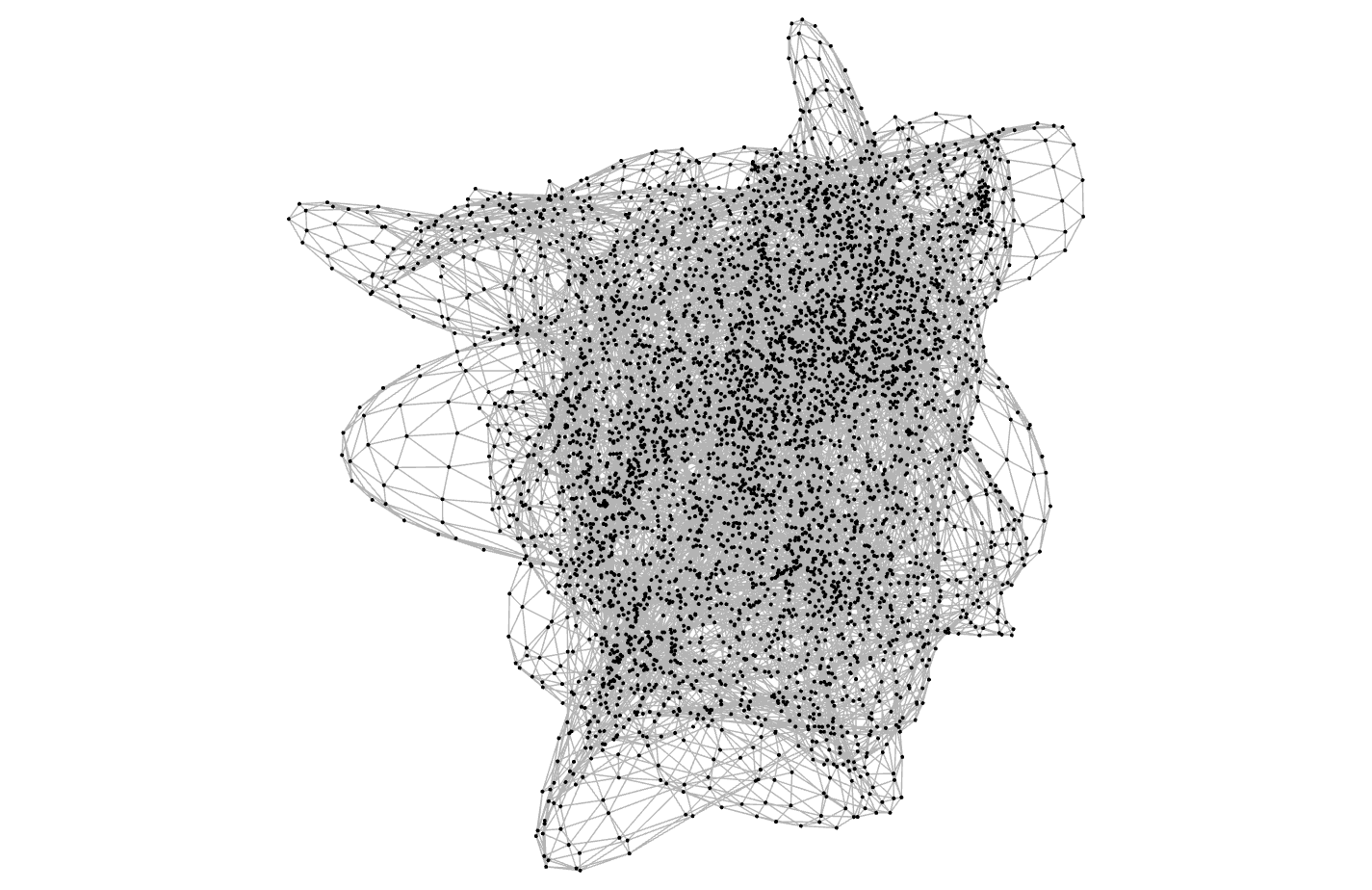}
  }
  \subfloat[FRG]{\label{fig:3elt_original_fre}%
    \centering
      \includegraphics[width=0.3\textwidth]{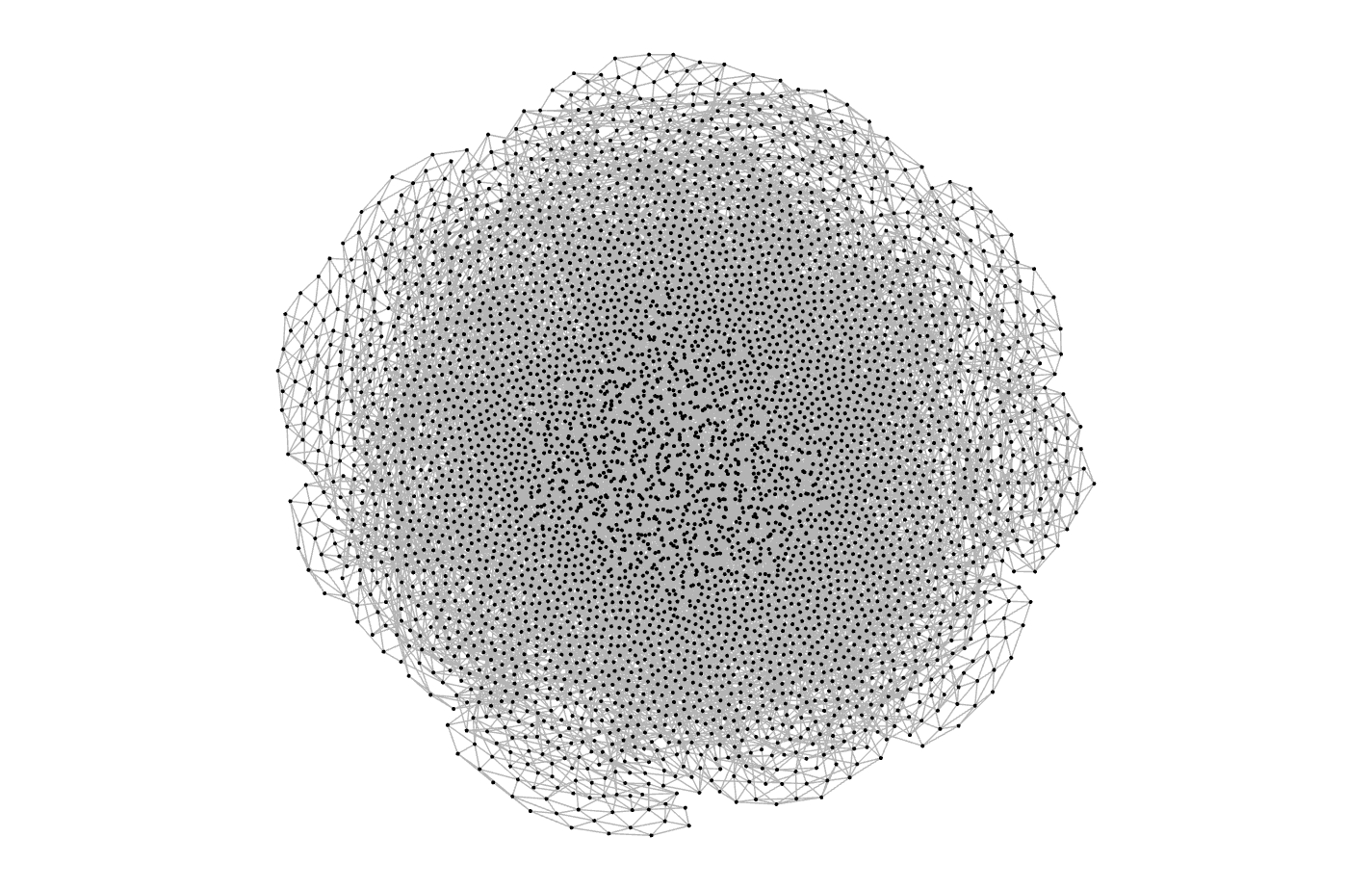}
   }
    \subfloat[FR]{\label{fig:3elt_original_frg}%
    \centering
      \includegraphics[width=0.3\textwidth]{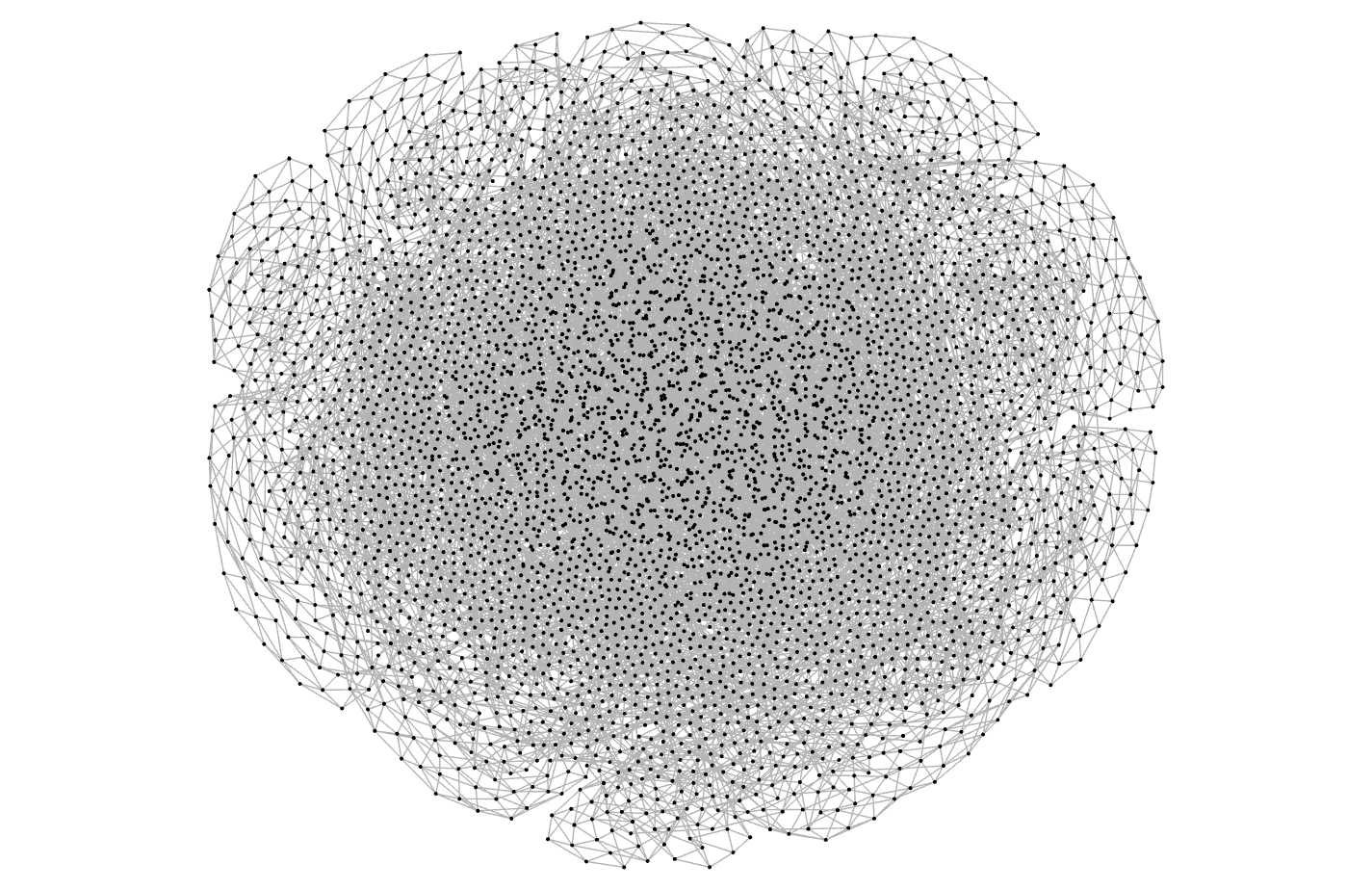}
   }
   \hfill
  \centering
  \subfloat[Infomap FME]{\label{fig:3elt_OGDF_ML_FME_infomap}%
    \centering
      \includegraphics[width=0.3\textwidth]{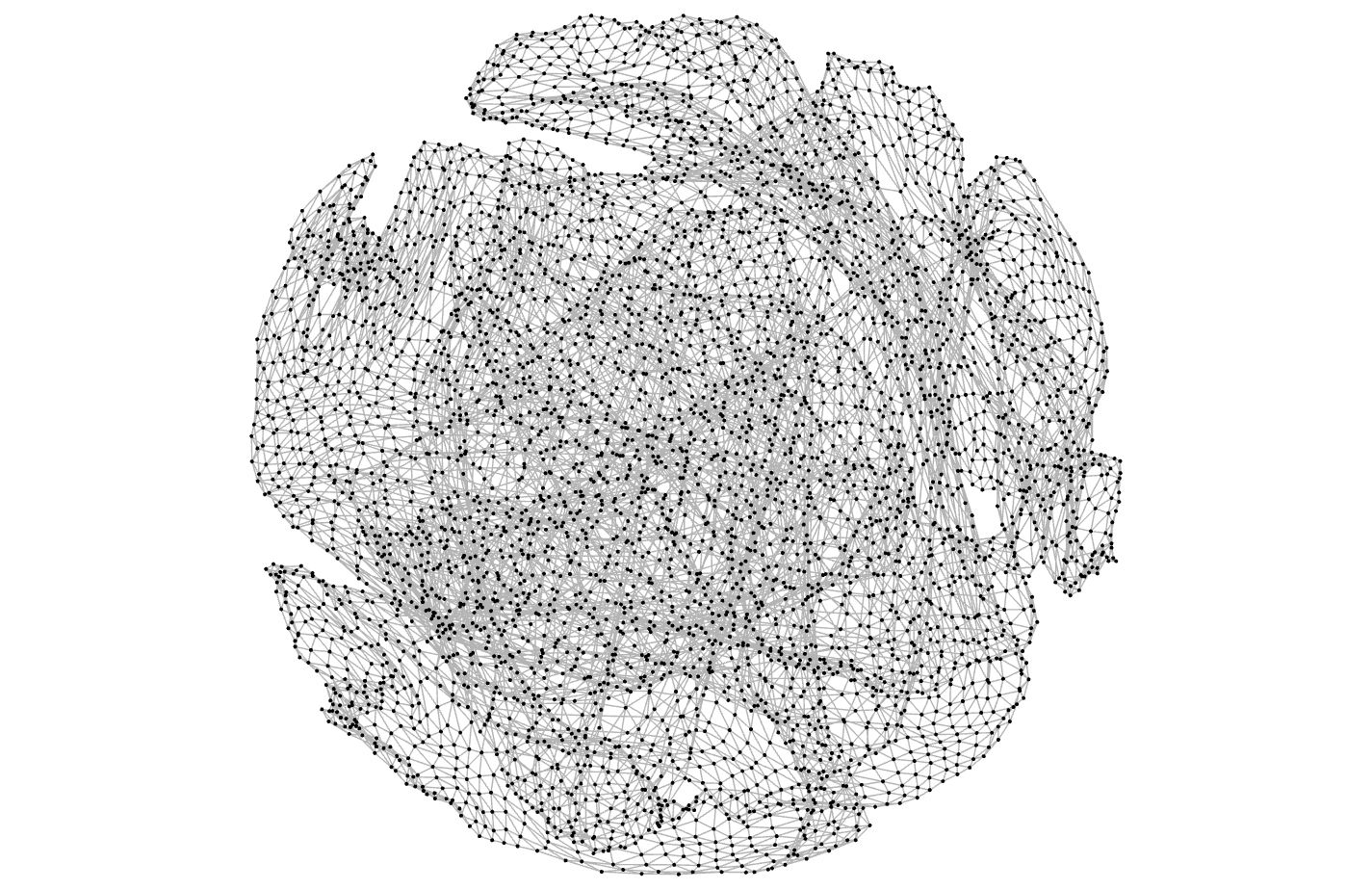}
  }
  \subfloat[Infomap FRG]{\label{fig:3elt_OGDF_ML_FR_infomap}%
    \centering
      \includegraphics[width=0.3\textwidth]{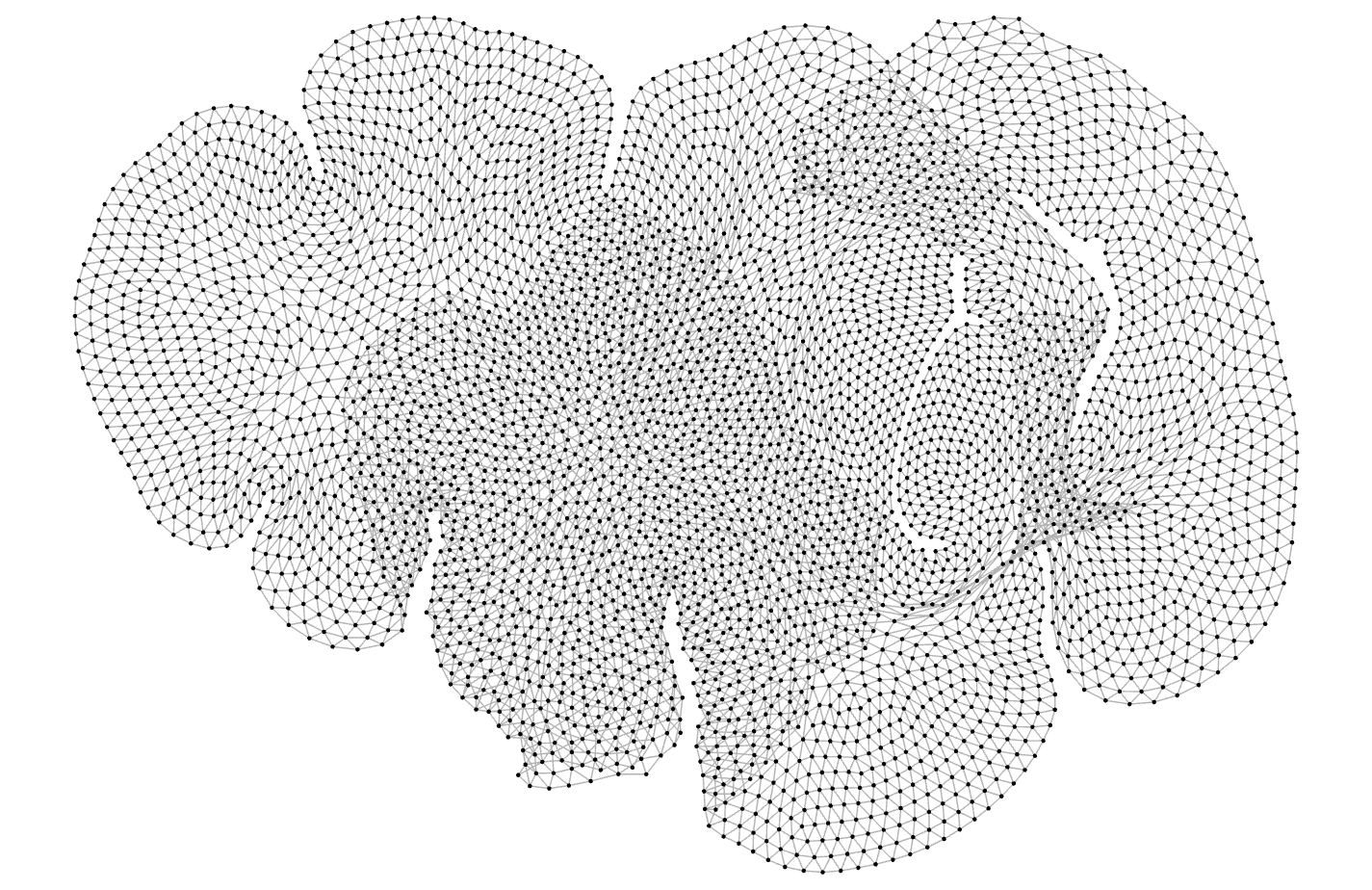}
  }
  \subfloat[Infomap FR]{\label{fig:3elt_OGDF_ML_FRE_infomap}%
    \centering
      \includegraphics[width=0.3\textwidth]{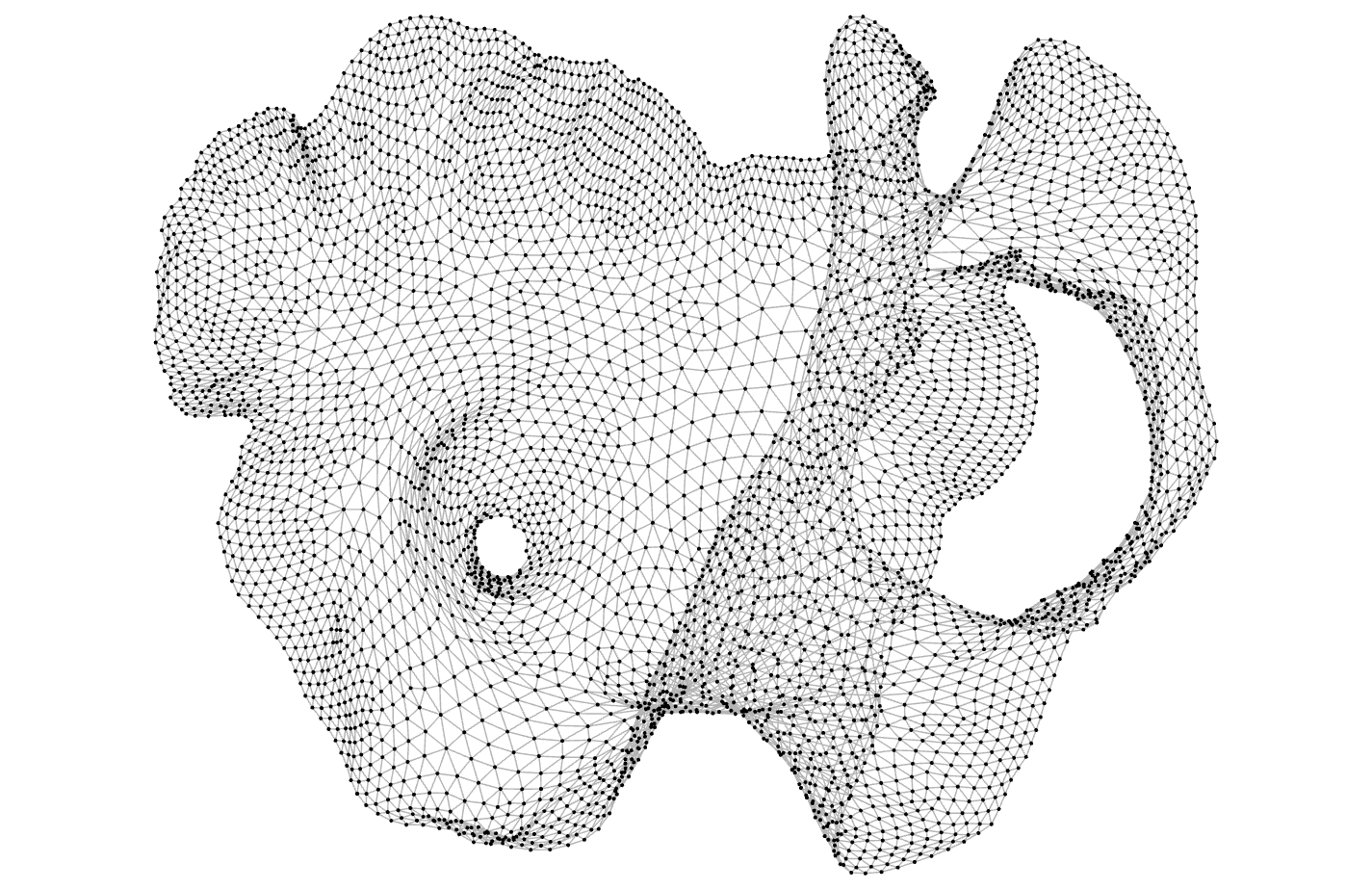}
  }
\caption{Visual comparison of $3elt$.}
\label{fig:3elt}
\end{figure}

\noindent{\bf Visual Comparison: }
Overall, Infomap multi-level layouts perform significantly better than original layouts.
In general, InfomapFR and InfomapFRG  perform significantly better than other layouts,
and InfomapFME achieved the most significant improvement over FME.
For example, Figure~\ref{fig:3elt} shows visual comparison between layouts of $3elt$.

\begin{figure}[h!]
  \centering
  \subfloat[$FM^3$]{\label{fig:USpowerGrid_original_fm3}%
    \centering
     \includegraphics[width=0.25\linewidth]{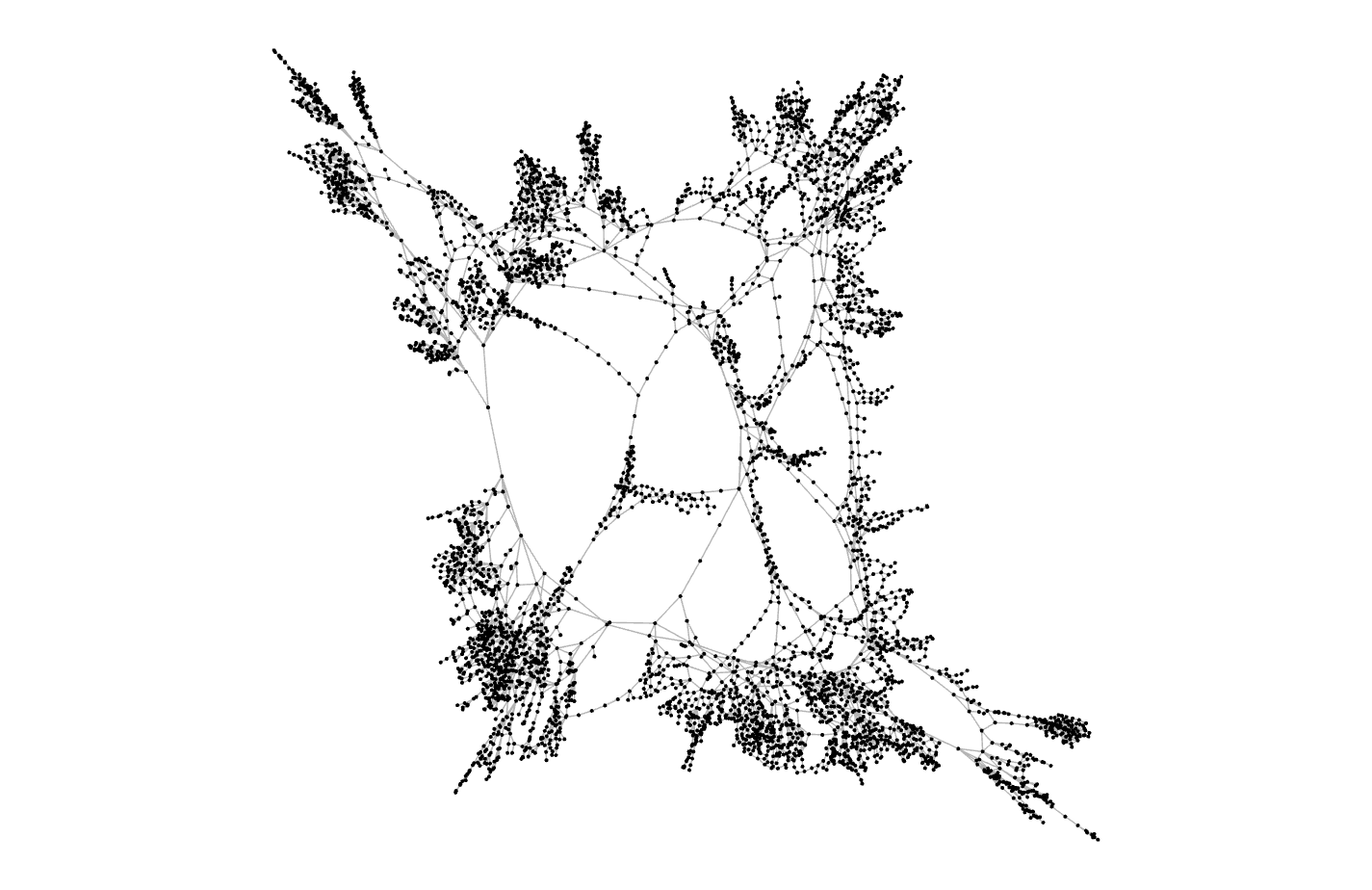}
  }
  \subfloat[Infomap FR]{\label{fig:USpowerGrid_OGDF_FRE_infomap}%
    \centering
      \includegraphics[width=0.25\linewidth]{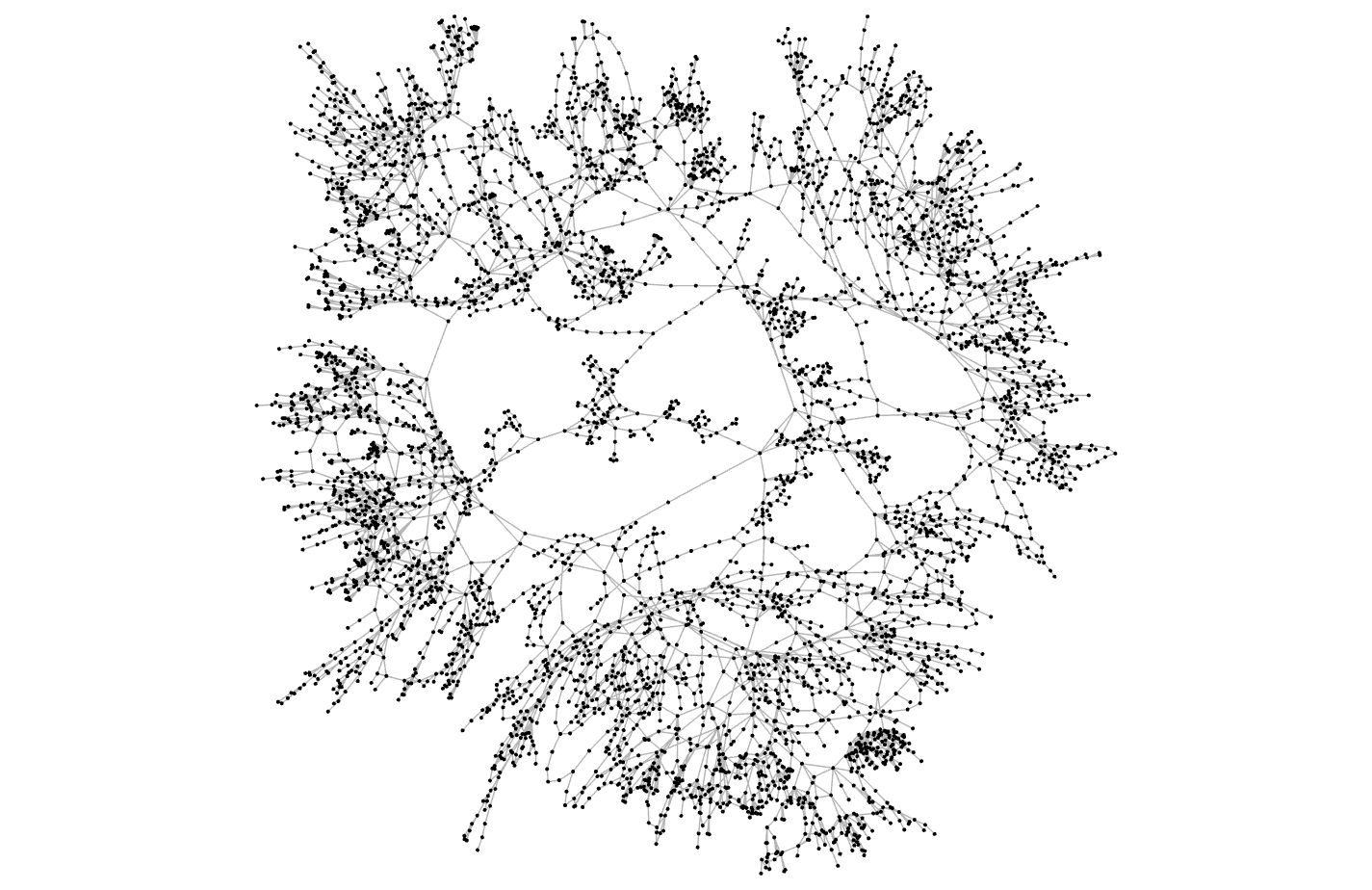}
   }
   \subfloat[$FM^3$]{\label{fig:shock-9_original_fm3}%
    \centering
     \includegraphics[width=0.25\linewidth]{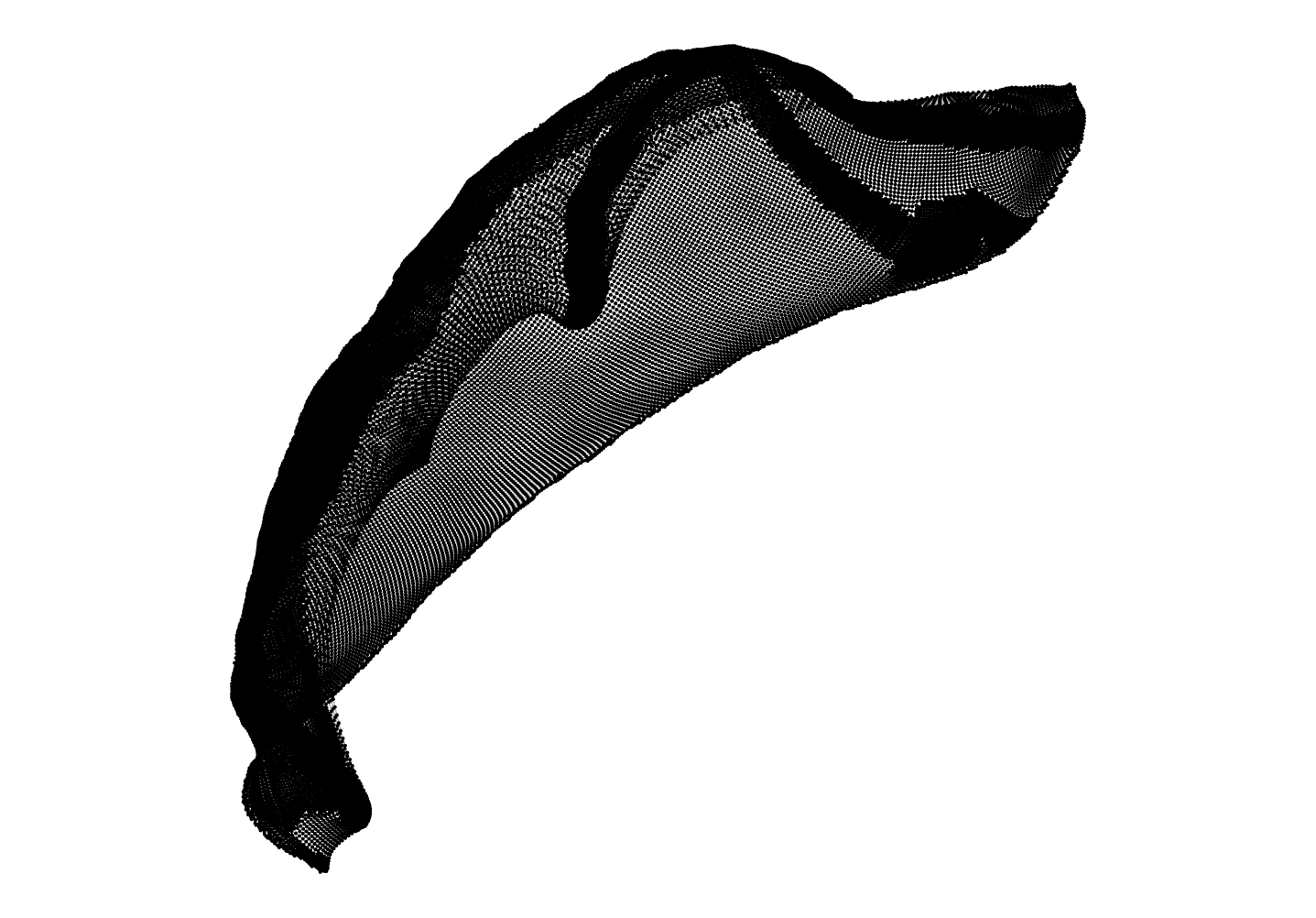}
  }
  \subfloat[Infomap FR]{\label{fig:shock-9_OGDF_FRE_infomap}%
    \centering
      \includegraphics[width=0.25\linewidth]{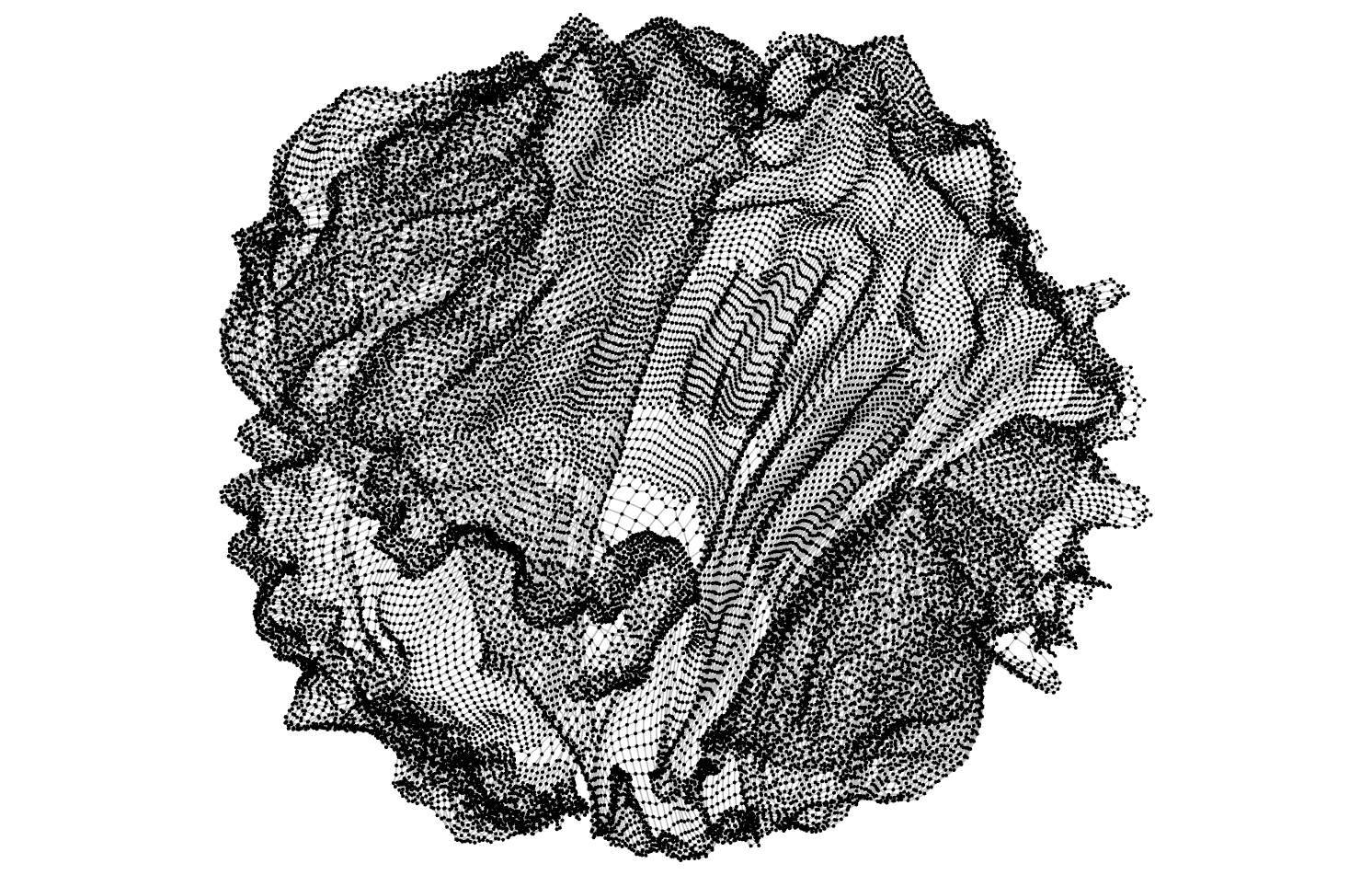}
   }
\caption{Comparison with $FM^3$: (a)(b)$USpowerGrid$; (c)(d) \emph{shock-9}}
\label{fig:FM3}
\end{figure}

\medskip
\noindent {\bf Comparison with $FM^3$:}
Figure~\ref{fig:average}(h) shows {\em average shape-based metrics} between InfomapFR, InfomapFRG and $FM^3$, excluding the outlier.
Clearly, we can see that InfomapFR and Infomap FRG perform similar to $FM^3$ in shape-based metrics.
For {\em layout comparison}, see Figures~\ref{fig:FM3}.
We can see that InfomapFR  perform similar to $FM^3$, and for some instances perform better than $FM^3$.

\medskip
\noindent {\bf Summary:}
Our experimental results provide strong evidence that our Infomap based multi-level algorithm performs considerably well
for real-world social networks, biological networks and benchmark graphs.

- Overall, Infomap multi-level layouts perform significantly better than
original layouts in terms of quality metrics and visualisation.

- Metric wise, InfomapFR and InfomapFRG  perform better than InfomapFME.

- InfomapFME achieved the most significant improvement.

- InfomapFR and InfomapFRG perform similar to $FM^3$ in terms of shape-based metrics and visual comparison.



\newpage
\bibliographystyle{splncs04}
\bibliography{gd}

\end{document}